\documentstyle[12pt]{article}

\newcommand{\be}{\begin{equation}}
\newcommand{\ee}{\end{equation}}
\newcommand{\prt}{\partial}
\newcommand{\al}{\alpha}
\newcommand{\vp}{\varphi}
\newcommand{\ep}{\varepsilon}
\newcommand{\sgm}{\sigma}
\newcommand{\dlt}{\delta}
\newcommand{\ra}{\rightarrow}

\begin{document}

\begin{center}
{\Large{\bf Stabilizing Role of Mesoscopic Fluctuations in 
Spin Systems} \\ [5mm]
V. I. Yukalov} \\ [2mm]
{\it Centre for Interdisciplinary Studies in Chemical Physics \\
University of Western Ontario, London, Ontario N6A 3K7, Canada \\
and \\
Bogolubov Laboratory of Theoretical Physics \\
Joint Institute for Nuclear Research, Dubna 141980, Russia}
\end{center}

\vspace{3cm}

\begin{abstract}

The occurrence of mesoscopic fluctuations in statistical systems implies,
from the point of view of dynamical theory, the existence of local
instabilities. However, the presence of such fluctuations can make a
system, as a whole, more stable from the thermodynamic point of view.
Thus, in many cases, a local dynamic instability is a requisite for the
global thermodynamic stability. This idea is illustrated by several spin
models.
\end{abstract}

\vspace{4cm}


\vspace{5mm}

{\bf keywords:} mesoscopic fluctuations, heterophase states, spin systems,
phase transitions

\newpage

\section{Introduction}

There are two general types of fluctuations in statistical systems,
microscopic and mesoscopic. The former are small oscillations about a
ground state, and they define collective excitations, such as phonons,
magnons, etc. This kind of fluctuations is pertinent to equilibrium state.
Collective excitations characterize a set of quantum states of a statistical 
system, and the word microscopic reflects the microscopic nature of such
fluctuations.

Contrary to these, {\it mesoscopic fluctuations} are such that make a
macroscopic system heterogeneous or locally drive a system out of 
equilibrium. Locally can mean in space, or in time, or both. So, mesoscopic 
fluctuations are, in general, nonequilibrium and are related to an averaged 
description of a statistical system. The word mesoscopic, in relation to
space, means that the characteristic size of such a fluctuation is much
larger than the average interparticle distance but is much smaller than
the size of a system itself. In relation to time, mesoscopic implies that
the characteristic lifetime of a mesoscopic fluctuation is much longer
than an effective oscillation period of microscopic fluctuations but much
shorter than the observation time. When mesoscopic fluctuations correspond
to the formation of nuclei of one thermodynamic phase inside another, they
are called {\it heterophase fluctuations} [1]. The latter are ubiquitous
in nature, and plenty of examples are described in review [1]. Recently,
much attention has been paid to the study of mesoscopic fluctuations in
high--temperature superconductors [2-7], where the corresponding
phenomenon is often termed phase separation [3-6], although it would be
more correct to call this effect {\it mesoscopic phase separation} in
order to distinguish it from the principally different Gibbs phase
separation occurring at macroscopic scales (see discussion in [1,7]).

When mesoscopic fluctuations are frozen in time, the corresponding system
looks like an ensemble of clusters with different properties. Such a
system can be even equilibrium [8,9]. And if these fluctuations are not
frozen in time, they make the system nonequilibrium [1]. The latter case
is more interesting than that of a frozen macroscopic structure, since it
involves three difficult questions: (i) How to develop a statistical
description of such a nonequilibrium and nonuniform system? (ii) Can this
mesoscopic state be an attractor and, if so, what kind of attractor is it?
(iii) Though the existence of nonequilibrium mesoscopic fluctuations
implies local instability, but is it possible that globally the system is
nevertheless stable?

The answer to the first question has been done by developing a consistent
statistical theory of systems with such mesoscopic fluctuations [1]. In
this approach, after averaging over these fluctuations, a renormalized
Hamiltonian is defined representing a set of phase replicas, each of which
describes an effective equilibrium system. To deal further with the
renormalized Hamiltonian, one may employ the techniques of equilibrium 
statistical mechanics.

As an answer to the second question, it has been conjectured that the state 
of a system with nonequilibrium mesoscopic fluctuations is a chaotic
attractor [1]. It was also shown [10] that a uniform statistical system is
structurally unstable with respect to arbitrary small random external
perturbations. One can recollect as well that in dynamical theory there
are plenty of examples of dynamical systems with chaotic attractors (see
e.g. [11-13] and references therein).

The third question has been considered for some simple spin models [1] and
analyzed in more detail for a chaotic lattice--gas model [14]. In the
present paper, the study of several other less trivial spin models is
given, with the aim to show that for each model there can be found a
region of parameters, where mesoscopic fluctuations do make the system
globally stable, that is, thermodynamically more stable than the analogous
system without these fluctuations.

\section{Effective Hamiltonian}

This section is a very brief recollection of the main definitions we shall
need in what follows for analyzing systems with mesoscopic or heterophase
fluctuations. We employ the theory of such systems developed in Ref. [1].
Throughout the text, the terms mesoscopic and heterophase will be used in
parallel.

Imagine that we are dealing with a system in which randomly in space and
time there arise mesoscopic fluctuations. For crystals and liquids, these
could be fluctuations of local space structure [15-18]. In the case of
spin systems, these are fluctuations of local magnetization [1]. Following
the general theory [1], we can average over such stochastic mesoscopic
fluctuations and obtain an averaged effective Hamiltonian
\be
H_{eff} =\oplus_\nu H_\nu ,
\ee
in which the index $\nu$ enumerates qualitatively different phases, and
$H_\nu$ is a phase--replica Hamiltonian representing a pure $\nu$--phase.
Each Hamiltonian $H_\nu$ is defined on a Hilbert space ${\cal H}_\nu$ of
microscopic states typical of the corresponding $\nu$--phase [1,19,20].
The averaged Hamiltonian (1) acts on a fiber space
\be
{\cal Y} =\otimes_\nu {\cal H}_\nu .
\ee
Hamiltonian (1) depends on a set $\{ w_\nu\}$ of geometric phase
probabilities $w_\nu$, with the properties
\be
\sum_\nu w_\nu = 1 , \qquad 0\leq w_\nu \leq 1 .
\ee
These phase probabilities are defined as quantities providing an absolute
minimum for the free energy
\be
f = -\frac{T}{N}\ln Tr\exp\left ( -\beta H_{eff}\right ) ,
\ee
where $T$ is temperature, $N$ is the averaged number of particles, 
$\beta T=1,\; k_B\equiv 1$, and the trace is taken over the fiber space
(2). The phase probabilities as functions of thermodynamic and Hamiltonian
parameters are given by the solutions of the equations
\be
\frac{\prt f}{\prt w_\nu} = 0 , \qquad
\frac{\prt^2f}{\prt w_\nu^2} > 0 ,
\ee
under the normalization condition (3).

As concrete examples, we analyse below spin systems. Heterophase states in
these systems are characterized by mesoscopic fluctuations of local
magnetization. Such spatial fluctuations can be observed in experiment by
means of neutron scattering methods, like diffuse scattering, small--angle
scattering, Bragg reflections, and polarized--beam scattering [21].

\section{Model with Competing Interactions}

Consider a one--dimensional Ising--type model modified by including two
kinds of spin interactions, long--range and short--range interactions,
\be
J_{ij} =\al I\dlt_{|i-j|,1} + ( 1 -\al )J^0_{ij} ,
\ee
where $J^0_{ij}$ satisfies the properties
$$ \lim_{N\ra\infty}J^0_{ij}=0 , \qquad 
\lim_{N\ra\infty}\frac{1}{N}\sum_{i\neq j}^NJ^0_{ij} = J < \infty $$
and $\al$ is a crossover parameter [22,23]. For $\al=0$, we have only a
long--range interaction, while for $\alpha=1$, we get the Ising
nearest--neighbour interactions. Assume that mesoscopic fluctuations are
heterophase fluctuations between two phases, ferromagnetic and paramagnetic. 
The averaged Hamiltonian (1) consists of the phase--replica terms
\be
H_\nu =w_\nu^2\left ( \frac{1}{2} NU -
\frac{1}{4}\sum_{i\neq j} J_{ij} s_is_j\right ) ,
\ee
where $N$ is the number of lattice cites, $U$ is a crystalline--field
parameter, and $s_i=\pm 1$. Note that the crystalline field cannot be
omitted since it influences the values of the phase probabilities,
although it does not contain spin variables [24]. This makes the situation
rather different from the case of pure monophase systems.

Each phase is characterized by an order parameter
\be
\sgm_\nu \equiv\frac{1}{N}\sum_{i=1}^N\langle s_i\rangle_\nu ,
\ee
where $\langle\ldots\rangle_\nu$ implies the statistical averaging over
the space of typical states ${\cal H}_\nu$; that is, an average
$\langle\hat A\rangle_\nu$ of an operator $\hat A$ means
$$ \langle \hat A\rangle_\nu \equiv Tr_\nu\rho_\nu\hat A , \qquad
\rho_\nu\equiv
\frac{\exp(-\beta H_\nu)}{Tr_\nu\exp(-\beta H_\nu)}\; , $$
where $Tr_\nu$ means a trace over ${\cal H}_\nu$. Let $\nu=1$ correspond
to ferromagnetic phase and $\nu=2$, to paramagnetic phase. Then, by
definition,
\be
\sgm_1\not\equiv 0 , \qquad \sgm_2\equiv 0 .
\ee
This condition relates the order parameters (8) with the space 
${\cal H}_\nu$ of typical states, making it possible to construct the
latter as quantum weighted spaces [1].

Calculating the specific free energy (4), we can use for the short--range
part of the Hamiltonian the transfer--matrix method (see e.g. [25]), and
the long--range part, as is known [26], is asymptotically equivalent to
the mean--field form. As a result, we obtain
$$ f = \left ( w^2 - w +\frac{1}{2}\right ) U - \frac{w^2}{4}\left [
\al T - ( 1 - \al ) J\sgm^2\right ] - $$
\be
- T\ln\left [ {\rm cosh}\vp +\sqrt{{\rm sinh}^2\vp +\exp(-4\vp_1)}\right ]
- T\ln\left ( 2{\rm cosh}\vp_2\right ) ,
\ee
where $w\equiv w_1,\;\sgm\equiv\sgm_1$, and
$$ \vp\equiv w^2\frac{(1-\al)J\sgm}{2T} , \qquad 
\vp_1\equiv w^2\frac{\al I}{4T} , \qquad 
\vp_2\equiv (1 - w)^2\frac{\al I}{4T} . $$
For what follows, it is convenient to introduce dimensionless quantities
\be
u \equiv\frac{U}{J} , \qquad g\equiv \frac{I}{J} , \qquad 
t\equiv\frac{T}{J} .
\ee
For the probability of the ferromagnetic phase, from the first of equations 
(5), we find the equation 
$$ \frac{4gw\al\exp(-4\vp_1)}{{\rm cosh}\vp\sqrt{{\rm sinh}^2\vp+
\exp(-4\vp_1)} +{\rm sinh}^2\vp +\exp(-4\vp_1)} + $$
\be
+ u(2w - 1) - 2\al g w - 2(1-\al )w\sgm^2 + 2\al (1 -w) g{\rm tanh}\vp_2
= 0 . 
\ee
For the order parameter $\sgm\equiv\sgm_1$, defined in (8), we get
\be
\sgm =\frac{{\rm sinh}\vp}{\sqrt{{\rm sinh}^2\vp +\exp(-4\vp_1)}} .
\ee
The heterophase state is thermodynamically more stable than the pure
state, if the inequality
\be
\Delta f\equiv f(1) - f(w) > 0 
\ee
holds true. When (5) is valid but (14) is not, the heterophase state is
metastable.

Let us investigate the stability of the system at $T=0$. Then, from (12),
we obtain the ferromagnetic--phase probability $w(T=0)\equiv w_0$ in the
form
\be
w_0 =\frac{2u-|\al|g}{4u - 1+\al -(\al +|\al|)g} \; .
\ee
Also, we have
$$ 
\frac{\prt^2f}{\prt w^2} = 2 ( 4u - \al g - 1 + \al -|\al|g ) , $$
$$ \frac{\Delta f}{J} =
\frac{(2u-\al g - 1+\al)^2(4u +2\al g - 1+\al)}{4(4u- 1 +\al)^2} , \qquad
\al < 0 , $$
$$ \frac{\Delta f}{J} =
\frac{(2u-\al g - 1 +\al)^2}{4(4u -2\al g - 1 +\al)} , \qquad
\al \geq 0 . $$

From conditions (3), (5), and (14), it follows that the heterophase state
is absolutely stable if either
\be
 u > \max\left\{ \frac{1}{4} ( 1 -\al - 2\al g) , \;
\frac{1}{2}( 1 -\al + \al g)\right \} , \qquad \al \leq 0 ,
\ee
or
\be
u > \frac{1}{2} ( 1 -\al +\al g) , \qquad \al \geq 0 .
\ee
The heterophase state is metastable if either
\be
\frac{1}{2} ( 1-\al +\al g ) < u < \frac{1}{4} ( 1-\al -2\al g) , \qquad
\al < 0 ,
\ee
or
\be
u <\frac{1}{2}\al g , \qquad \al \geq 0 .
\ee
In the case when
\be
\frac{1}{2}|\al|g < u <\frac{1}{2} ( 1 -\al +\al g) ,
\ee
the system at zero temperature is purely ferromagnetic, but beginning from
a finite temperature $T_n$, called the nucleation temperature [1] and
defined by the condition $w(T_n)=1$, the heterophase state becomes
profitable, being a mixture of ferromagnetic and paramagnetic phases.

At low temperatures, when $t\ra 0$, we find the following asymptotic
behaviour for the ferrophase probability
$$ \frac{w}{w_0} \simeq 1 - 
\frac{4(1-\al +\al g)}{4u-\al g - 1+\al -|\al|g}\exp\left\{ -
\frac{w_0^2}{t} ( 1 -\al +4\al g)\right\} + $$
$$ + \frac{2|\al|g(2u-\al g - 1 +\al)}{(2u-|\al|g)(4u-\al g-1+\al-|\al|g)}
\exp\left\{ -\frac{(1-w_0)^2}{t}2|\al|g\right \} , $$
the order parameter
$$ \sgm\simeq 
1 - 2\exp\left\{ -\frac{w_0^2}{t} (1 -\al + 4\al g)\right\} , $$
the entropy
$$ S\simeq \frac{w_0^2}{t} ( 1-\al +\al g)\exp\left\{ -
\frac{w_0^2}{t} ( 1 -\al +4\al g)\right\} + $$
$$ +\frac{(1-w_0)^2}{2t}|\al|g\exp\left\{ -
\frac{(1-w_0)^2}{t}2|\al|g\right\} , $$
and for the heat capacity
$$ C_V\simeq \frac{w_0^4}{t^2} (1 -\al +\al g)^2\exp\left\{ -
\frac{w_0^2}{t} ( 1 -\al +4\al g)\right\} + $$
$$ +\frac{(1-w_0)^4}{4t^2}\al^2g^2\exp\left\{ -
\frac{(1-w_0)^2}{t}2|\al|g\right\} . $$
The positivity of the specific heat indicates that the heterophase state
is stable with respect to thermal fluctuations.

Now, let us analyse the critical behaviour of the model. The critical
temperature $t_c$ is defined by the condition $\sgm(t_c)=0$, which gives
\be
t_c =\frac{1-\al}{8}\exp\left ( \frac{\al g}{8t_c}\right ) .
\ee
As follows from (21), there exists a negative value of the crossover
parameter $\al=\al_0$,
\be
\al_o = -\frac{1}{eg -1} \qquad (eg > 1) ,
\ee
such that for $\al<\al_0$ the ferromagnetic state is impossible at all
temperatures. But if $g\leq e^{-1}=0.3679$, then a positive solution for
$t_c$ is available for any $\al< 1$. The appearance of the limiting value
(22) is quite explicable. Really, negative values of $\al$ correspond to
the antiferromagnetic character of the short--range interaction. The
presence of an interaction having the opposite sign, as compared to the
ferromagnetic long--range interaction, serves  as a disordering factor.
The onset of ferromagnetic order is possible only it the disordering
short--range interaction is not too large. One might recollect several
other examples when an ordering in a system occurs only if some limiting
relations between competing interactions take place. Recall, for instance,
the criteria of magnetism in the Hubbard model [27,28].

It is interesting that the crossover behaviour of the critical
temperature, between the mean--field value $t_c=\frac{1}{8}\; (\al=0)$ and
the short--range case $t_c=0\; (\al=1)$, is nonmonotonic. The maximum of
(21) occurs at
\be
t_{max} =\frac{g}{8(1+\ln g)} , \qquad \al_{max} =\frac{\ln g}{1+\ln g} .
\ee
The ratio of this maximum to the mean--field critical temperature
$t_c=1/8$, that is,
$$ 8t_{max} =\frac{g}{1+\ln g} , $$
can become arbitrary large for  $g\gg 1$.

The presence of mesoscopic fluctuations, as well as the antiferromagnetic
short--range interaction, can change the second--order phase transition to
the first--order one. The region of first--order phase transition is
defined either by the inequalities
\be
u_0 < u < u_t \qquad (\al >\al_1) ,
\ee
or by the inequalities 
\be
u_0 < u < u_0 - |u_t - u_0 | \qquad (\al_0 <\al <\al_1 ) ,
\ee
where
$$ u_t = u_0 +\frac{12(1-\al)(\al g + 2t_c)^2}{12t_c^2 - (1-\al)^2} , $$
\be
u_0 \equiv\frac{4\al(1-\al)g + 4t_c^2 -(1-\al)^2}{4(1+t_c-\al)^2}\; ,
\ee
$$ \al_1 = 1 -2\sqrt{3} t_c \; . $$

The value $u=u_t$, defined in (26), is a function of the parameters $\al$
and $g$. The equation $u=u_t(\al,g)$ describes a surface on which the
order of phase transition changes. This is called a tricritical surface.
On the latter, the critical indices also change by a jump. Thus,
considering the asymptotic behaviour of the specific heat 
$C_V\propto |\tau|^{-\al}$, order parameter $\sgm\propto |\tau|^\beta$, and
susceptibility $\chi\propto |\tau|^{-\gamma}$, when approaching the critical
point $t_c$, so that $\tau\equiv (t-t_c)/t_c\ra -0$, we obtain
\be
\al=0, \qquad \beta=\frac{1}{2} , \qquad \gamma=1 \qquad (u\neq u_t)
\ee
outside the tricritical surface, and
\be
\al =\frac{1}{2} , \qquad \beta=\frac{1}{4} , \qquad \gamma=1 \qquad (u=u_t)
\ee
on the tricritical surface.

One more critical index can be introduced [1] for the phase probability
$w$ as
\be
w-\frac{1}{2} \propto |\tau|^\ep .
\ee
This index $\ep$ is specific for heterophase systems. In our case we find
that this index also jumps on the tricritical surface $u=u_t(\al,g)$,
\begin{eqnarray}
\ep=\left\{ \begin{array}{cc}
1,           & u\neq u_t \\
\frac{1}{2}, & u=u_t .
\end{array}\right.
\end{eqnarray}

Despite the seeming simplicity of the model considered, it displays quite
nontrivial behaviour. One of the most interesting features is a strongly
nonmonotonic dependence of the critical temperature on the crossover
parameter $\al$. Also, in the space of three parameters, $u,\;\al$, and
$g$, there is a region, where mesoscopic fluctuations make the system
thermodynamically more stable than that system without such fluctuations.

\section{Heterophase Spin Glass}

Consider a generalization of the Sherrington--Kirkpatrick spin--glass
model [29] to the case of a system with mesoscopic fluctuations. We shall
keep in mind paramagnetic fluctuations inside the spin--glass phase [30].
Let interactions $J_{ij}$ between spins at sites $i,j=1,2,\ldots,N$ be
distributed by the Gaussian law
\be
P(J_{ij}) =\frac{1}{J}\left (\frac{N}{2\pi}\right )^{1/2}\exp\left\{ -
\frac{N}{2J^2}\left ( J_{ij} -\frac{J_0}{N}\right )^2\right\} .
\ee
For simplicity, we put $J_0\equiv 0$. The average over interactions for a
function $A\{ J_{ij}\}$ of a set $\{ J_{ij}\}$ of interactions $J_{ij}$ is
defined as
\be
[ A\{ J_{ij}\} ]_{av} \equiv
\int A\{ J_{ij}\}\prod_{i\neq j}P(J_{ij})dJ_{ij}\; .
\ee

For an Ising--like system with mesoscopic fluctuations, following the
general renormalization procedure [1], we have an effective Hamiltonian
(1) with the phase--replica terms
\be
H_\nu\{ J_{ij}\} = w_\nu^2\left ( \frac{1}{2} NU -
\sum_{i\neq j} J_{ij} s_i s_j\right ) ,
\ee
in which $U$ is a crystalline--field constant and $s_i=\pm 1$. To
distinguish phases, we need an order parameter. For spin glasses, this is
the Edwards--Anderson [31] order parameter
\be
q_\nu \equiv\left [\langle s_i\rangle_\nu^2\right ]_{av}\;  ,
\ee
where $\langle\ldots\rangle_\nu$ means a statistical averaging with the
Hamiltonian (33) over the space ${\cal H}_\nu$ of states typical of a
phase with the parameter (34), and $[\ldots]_{av}$ denotes the averaging
(32) over interactions. Let the spin--glass phase be indexed by $\nu=1$
and the paramagnetic phase, by $\nu=2$. Then, by definition,
\be
q_1\not\equiv 0 , \qquad q_2\equiv 0 .
\ee
The ferromagnetic phase, because of $J_0\equiv 0$, is absent.

The free energy (4) writes
\be
f = f_1 + f_2 , \qquad f_\nu = \left [ f_\nu\{ J_{ij}\}\right ]_{av} \; ,
\ee
where
$$ f_\nu\{ J_{ij}\} =-\frac{T}{N}\ln Tr_\nu\exp\left ( -\beta H_\nu\{
J_{ij}\}
\right ) , $$
and ${\cal H}_\nu=\{s_i|\; i=1,2,\ldots,N;q_\nu\}$ is the space of
states typical of a phase with the order parameter (34). With the replica
trick [31], one has
$$ \left [\ln Z\{ J_{ij}\}\right ]_{av} =\lim_{n\ra 0}\frac{1}{n}\left (
\left [ Z^n\{ J_{ij}\}\right ]_{av} - 1\right ) =
\lim_{n\ra 0}\frac{\prt}{\prt n}\left [ Z^n\{ J_{ij}\}\right ]_{av}\; . $$
Using this, for the free energy (36), we obtain
$$ f =\frac{U}{2} \left [ w^2 +( 1 - w)^2\right ] - 
\frac{1}{4}\beta J^2 w^4 (1 - q)^2 - \frac{1}{4}\beta J^2 ( 1 - w)^4 - $$
\be
- T\int_{-\infty}^{+\infty} 
p(x)\ln\left [ 2{\rm cosh}(\beta Jw^2q^{1/2}x)\right ] dx - T\ln 2 ,
\ee
where $w\equiv w_1,\; q\equiv q_1$, and
$$ p(x) =\frac{1}{\sqrt{2\pi}}\exp\left (-\frac{1}{2}x^2\right ) . $$
For the spin--glass order parameter (34), with $\nu=1$, we get
\be
q =\int_{-\infty}^{+\infty} 
p(x){\rm tanh}^2\left (\beta Jw^2q^{1/2} x\right ) dx .
\ee
The trivial solution $q=0$ of (38) is excluded by condition (35). From the
first of eqs. (5), we get an equation
\be
w^3 (1 - q)^2 - ( 1 - w)^3 - u(2w - 1 )t = 0
\ee
for the probability $w$ of the spin--glass phase, where
\be
u\equiv\frac{U}{J}\; , \qquad t\equiv\frac{T}{J}\; .
\ee
Among three roots of eq. (39), we have to choose that one satisfying the
normalization condition (3). In comparison to the standard spin glass
[31,32], the phase probability $w$ plays the role of an additional order
parameter.

Let us analyse the thermodynamic characteristics of the heterophase spin
glass. It is convenient to introduce the notation
\be
u_0\equiv2\sqrt{\frac{2}{\pi}} = 1.595769 .
\ee
At low temperatures, the spin--glass order parameter behaves  as
$$ q\simeq 1 -\frac{u_0}{2} t - u_0(u_0 - u)t^{4/3} \qquad (u< u_0) , $$
\be
q\simeq 1 -\frac{u_0}{2} t -\frac{1}{\pi} t^2 \qquad (u\geq u_0) ,
\ee
where $t\ra 0$. The asymptotic, as $t\ra 0$, behaviour of the phase
probability is
$$ w\simeq 1 - (u_0 - u) t^{1/3} \qquad ( u < u_0 ) , $$
\be
w\simeq 1 \qquad (u\geq u_0) .
\ee
For the specific heat and entropy, we find
$$ C_V \simeq \frac{1}{6} ( u_0 - u )^{4/3} t^{-2/3} \qquad (u<u_0) , $$
\be
C_V \simeq \frac{(\pi^3 - 6)}{24\pi}u_0 t \qquad (u\geq u_0) ,
\ee
as $t\ra 0$, and, respectively,
$$ S\simeq -\frac{1}{4} (u-u_0)^{4/3} t^{-2/3} \qquad (u<u_0) , $$
\be
S\simeq \ln 2 -\frac{1}{2\pi} = 0.53399 \qquad (u\geq u_0) .
\ee
As follows from these expressions, the ground state is a pure spin--glass
phase: $w=1$ at $t=0$.

When the crystal--field parameter $u<u_0$, the system is unstable at low
temperatures, since $C_V\ra\infty$ and $S\ra -\infty$, in analogy
with the Sherrington--Kirkpartick case [29]. But for $u\geq u_0$,
mesoscopic paramagnetic fluctuations stabilize the system making the
behaviour of the specific heat and entropy normal.

In the vicinity of the critical point
\be
t_c =\frac{1}{4} \qquad (q=0) ,
\ee
we have for the spin--glass parameter
\be
q\simeq |\tau| \qquad \left ( \tau \equiv\frac{t-t_c}{t_c}\ra - 0\right )
\ee
and for the phase probability
\be
w\simeq \frac{1}{2} -\frac{\tau^2}{4(u-3)}\; .
\ee
From the stability condition
\be
\frac{\prt^2f}{\prt w^2}\simeq 2J (u - 3) > 0 \qquad (t\ra t_c) ,
\ee
we conclude that the second--order transition between the spin--glass and
paramagnetic phases occurs if $u>3$. The value $u=u_t=3$ corresponds to a
tricritical point, where the second--order transition changes for the
first order transition being realized for $u<3$. The critical index for
the phase probability, defined in (29), is
\be
\ep\equiv\lim_{\tau\ra 0}
\frac{\ln\left| w-\frac{1}{2}\right |}{\ln|\tau|} = 2 \qquad (u<3) .
\ee

In this way, mesoscopic paramagnetic fluctuations, when $u>u_0$, stabilize
the Sherrington--Kirkpatrick mean--field spin glass, making its specific
heat finite and entropy positive. In order to check whether the
heterophase spin glass becomes absolutely stable, one has to consider as
well the sign of magnetic susceptibility. According to our analysis [30],
the latter is positive, at least in the critical region. The mean--field
glass, as is known, can be made stable by invoking, for the
Edwards--Anderson order parameter, solutions with a broken replica
symmetry [33]. Generalizing this type of spin glass to the case including
mesoscopic fluctuations, it is possible to show [30] that, again, for
sufficiently large crystal--field parameter $u$, the free energy of the
heterophase spin glass becomes lower than of a pure spin glass. That is,
mesoscopic fluctuations can make the spin--glass system thermodynamically
more stable.

\section{Systems with Magnetic Reorientations}

The appearance of coexisting magnetic phases with different directions of
magnetization is characteristic of spin--reorientational transitions in
small or zero external magnetic fields. The existence of such mixed states
is well documented by a large number of experiments and have been
discussed in detail in books [34,35] and reviews [1,36,37]. The standard
theoretical description of magnetic reorientations, going through
intermediate mixed states, is done by means of Landau expansions involving
a set of fitting functions taken from experimental data [34,35]. This is
a purely phenomenological treatment giving no physical insight. Another
approach to systems with magnetic reorientations can be based on the
theory of mesoscopic fluctuations [1], with taking account of phase
fluctuations corresponding to different angle phases with mutually
orthogonal magnetizations [24,38,39]. Following such a microscopic
approach, it is possible to show that the balance between phase
probabilities and, as a result, magnetic reorientations are governed by
the tendency of a system to reach the state of an absolute thermodynamic
stability by allowing the appearance of mesoscopic fluctuations.

Consider a system in which there can coexist four different phases, three
of them, magnetic, having orthogonal to each other nonzero magnetizations
and one, paramagnetic, with zero magnetization. These phases will be
enumerated by the index $\nu=1,2,3,4$, and the notation $\{\nu\}=\{\al,4\}$, 
where $\al=1,2,3$, will be used. To separate the phases, we need an order
parameter, whose definition, as usual for spin systems, is based on the
average spin operator
\be
\stackrel{\ra}{S} \equiv\frac{1}{N}\sum_{i=1}^N\stackrel{\ra}{S}_i =
\left\{ S^\al\right\} ,
\ee
in which $\al=1,2,3$ and $\stackrel{\ra}{S}_i$ is a spin operator at a
lattice site $i=1,2,\ldots,N$. The vector order parameter of a
$\nu$--phase is defined as
\be
\stackrel{\ra}{\eta}_\nu \equiv\langle\stackrel{\ra}{S}\rangle_\nu =
\{\eta_\nu^\al\} ,
\ee
so that
\be
\eta_\al^\beta =\delta_{\al\beta}\eta_\al\; , \qquad \eta_4^\al\equiv 0 ,
\ee
where
\be
\eta_\al\equiv\langle S^\al\rangle_\al \not\equiv 0 .
\ee
In another way, we could write
\be
\stackrel{\ra}{\eta}_\al =\eta_\al\stackrel{\ra}{e}_\al , \qquad
\stackrel{\ra}{\eta}_4\equiv 0 ,
\ee
where $\eta_\al$ is given by (54) and $\stackrel{\ra}{e}_\al$ is a unit
vector along the $\al$--axis. The order parameters (55) define the
directions of magnetization for the related phases, three of which are
magnetic and one is paramagnetic.

Following the general theory [1] for a spin system with anisotropic
interactions $J_{ij}^\al$, after averaging over random phase
configurations, we come to an effective Hamiltonian (1) with the
phase--replica terms
\be
H_\nu = Nw_\nu K + w_\nu^2\left ( NU -
\sum_{i\neq j}^N\sum_{\al=1}^3  J_{ij}^\al S_i^\al S_j^\al\right ) ,
\ee
where $K$ is a mean kinetic energy per site for electrons and ions. Recall
[24] that the term $NU$ is the total potential energy, not including spin
operators, of electrons and ions in a crystalline lattice. Therefore, the
parameter $U$ can be called the crystal--field parameter, structural
constant, configurational energy per site, or lattice energy per site. In
general, the values $K,\; U$ and $J_{ij}^\al$ can also depend on the type
of a phase, but, for simplicity, we assume that they are the same for all
thermodynamic phases.

Minimizing the free energy (4), we get an equation
\be
w_\nu =\left ( \sum_{\mu=1}^4\frac{U-B_\nu}{U-B_\mu}\right )^{-1}
\ee
for the phase probabilities $w_\nu$, where
\be
B_\nu\equiv\frac{1}{N}\sum_{i\neq j}^N\sum_{\al=1}^3J_{ij}^\al\langle
S_i^\al S_j^\al\rangle_\nu .
\ee
From the second of eqs. (5), we have the stability condition
\be
U >\frac{1}{2}\left ( \sup_{\al}\{ B_\al\} + B_4\right ) ,
\ee
while the inequality $0<w_\nu<1$ is valid when
\be
U >\sup_{\al}\{ B_\al\} .
\ee
For ferromagnets, the interactions $J_{ij}^\al$ are positive, thence
$B_\nu>0$, and for antiferromagnets, $J_{i\al}^\al$ are negative, hence
$B_\nu<0$. Therefore, heterophase fluctuations appear easier in
antiferromagnets than in ferromagnets, as follows from (59) and (60).

To proceed further, we need to invoke some approximation. In what follows,
we use the mean--field decoupling
$$ \langle S_i^\al S_j^\al\rangle_\nu =\langle S_i^\al\rangle_\nu
\langle S_j^\al\rangle_\nu , $$
which yields
$$ B_\al=J_\al S^2\eta_\al^2, \quad B_4=0; \qquad 
J_\al\equiv\frac{1}{N}\sum_{i\neq j}^N J_{ij}^\al , $$
where $S$ is a spin value and $\eta_\al$ is given by (54). For $S=1/2$,
we have
\be
\eta_\al =\frac{1}{2}{\rm tanh}(\beta w_\al^2J_\al\eta_\al ) , \quad
\eta_4\equiv 0 .
\ee
The free energy (4) becomes
\be
f=\sum_{\al=1}^3\left\{ w_\al^2 ( U + J_\al\eta_\al^2 ) - T\ln [ 2
{\rm cosh}(\beta w_\al^2J_\al\eta_\al ) ]\right\} + w_4^2U - T\ln 2 .
\ee

In this way, the thermodynamic behaviour of the system is defined by the
set of seven coupled equations: four equations for the phase probabilities
(57) and three nontrivial equations for the order parameter (61). Among
all admissible solutions, one has to choose those satisfying all stability
conditions and providing an absolute minimum of the free energy (62). In
addition to the case when the phase probabilities are found from (57), we
need to consider the cases when one or several phase probabilities are put
zero [38]. This implies that we have to compare fifteen types of
solutions:

\vspace{2mm}

(1) $w_1\not\equiv 0, \quad w_2\not\equiv 0, \quad w_3\not\equiv 0, \quad
w_4\not\equiv 0;$
\vspace{2mm}

(2) $w_1\equiv 0, \quad w_2\not\equiv 0, \quad w_3\not\equiv 0, \quad
w_4\not\equiv 0;$

\vspace{2mm}

(3) $w_1\not\equiv 0, \quad w_2\equiv 0, \quad w_3\not\equiv 0, \quad
w_4\not\equiv 0;$

\vspace{2mm}

(4) $w_1\not\equiv 0, \quad w_2\not\equiv 0, \quad w_3\equiv 0, \quad
w_4\not\equiv 0;$

\vspace{2mm}

(5) $w_1\equiv 0, \quad w_2\equiv 0, \quad w_3\not\equiv 0, \quad
w_4\not\equiv 0;$

\vspace{2mm}

(6) $ w_1\equiv 0, \quad w_2\not\equiv 0, \quad w_3\equiv 0, \quad
w_4\not\equiv 0;$

\vspace{2mm}

(7) $w_1\not\equiv 0, \quad w_2\equiv 0, \quad w_3\equiv 0, \quad
w_4\not\equiv 0;$

\vspace{2mm}

(8) $w_1\equiv 0, \quad w_2\equiv 0, \quad w_3\equiv 0, \quad
w_4\equiv 1;$

\vspace{2mm}

(9) $w_1\not\equiv 0, \quad w_2\not\equiv 0, \quad w_3\not\equiv 0, \quad
w_4\equiv 0;$

\vspace{2mm}

(10) $w_1\equiv 0, \quad w_2\not\equiv 0, \quad w_3\not\equiv 0, \quad
w_4\equiv 0;$

\vspace{2mm}

(11) $w_1\not\equiv 0, \quad w_2\equiv 0, \quad w_3\not\equiv 0, \quad
w_4\equiv 0;$

\vspace{2mm}

(12) $w_1\not\equiv 0, \quad w_2\not\equiv 0, \quad w_3\equiv 0, \quad
w_4\equiv 0;$

\vspace{2mm}

(13) $w_1\equiv 1, \quad w_2\equiv 0, \quad w_3\equiv 0, \quad
w_4\equiv 0;$

\vspace{2mm}

(14) $w_1\equiv 0, \quad w_2\equiv 1, \quad w_3\equiv 0, \quad
w_4\equiv 0;$

\vspace{2mm}

(15) $w_1\equiv 0, \quad w_2\equiv 0, \quad w_3\equiv 1, \quad
w_4\equiv 0 .$

\vspace{2mm}
Leaving aside thermodynamics of this model [38,39], we would like to
concentrate here on the fact that choosing the most stable solutions makes
it possible to describe various reorientation transitions.

Let us arrange the exchange integrals in the order of their magnitude so
that
$$ 0 < J_1 < J_2 < J_3 . $$
The reorientation temperatures are defined by the equations
\be
\eta_\al (T_\al ) = 0 \qquad (\al=1,2,3 ) .
\ee
The largest reorientation temperature is the critical temperature
\be
T_c\equiv \sup_\al T_\al
\ee
for a ferromagnet--paramagnet phase transition. When this transition is of
first oreder, the transition temperature will be denoted by $T_0$. The
temperature at which a pure state transforms into a mixture is called [1]
the nucleation temperature. The latter is defined as
\be
T_n\equiv \inf_\nu T_n^\nu , \qquad w_\nu(T_n^\nu) = 0 ,
\ee
where $\nu=1,2,3,4$.

The arising sequence of phase transitions can be classified according to
the value of the parameter $U$ corresponding to the potential lattice
energy per site. When $U<0$, no heterophase states appear, and no
reorientation transitions occur. The sole transition is the 
ferromagnet--paramagnet phase transition at $T_c=\frac{1}{2}J_3$ to which
the notation
$$ [0\; 0\; 1]\leftarrow 2,T_c \rightarrow [0\; 0\; 0] $$
can be ascribed, where the numbers inside the square brackets means the
existence of a nonzero or zero magnetization along an axis $\al=1,2,3$,
respectively, and the number between the arrows shows the phase--transition 
order at a temperature $T_c$.

When $0<U\leq U_0$, where the value $U_0$ depends on magnitudes of $J_\al$,
then there appears the sequence of phase transitions
$$ [0\;0\; 1]\leftarrow 1,T_n\rightarrow [1\; 1\; 1]\leftarrow 1,T_0
\rightarrow [0\; 0\; 0] , $$
in which $T_n=T_1=T_2$. Both the nucleation and ferromagnet--paramagnet
transitions are of first order.

Increasing further the lattice--energy parameter, in an interval
$U_0<U\leq U_1$, we get the sequence
$$ [0\;0\; 1]\leftarrow 1,T_n\rightarrow [1\; 0\;1]\leftarrow 2,T_2
\rightarrow [1\; 1\; 1]\leftarrow 1,T_0\rightarrow [0\;0\;0] $$
of phase transitions, with $T_n=T_1$. At the value $U=U_1$, the nucleation
temperature $T_n$ corresponds to a tricritical point.

For $U_1<U\leq U_2$, we have
$$ [0\;0\;1]\leftarrow 2,T_n\rightarrow [0\;0\;1]\leftarrow2,T_1\rightarrow
[1\;0\;1]\leftarrow2,T_2\rightarrow [1\;1\;1]\leftarrow 1,T_0\rightarrow
[0\;0\;0] .$$
The nucleation becomes a second--order transition.

When $U_2<U\leq U_3$, then the sequence of phase transitions simplifies to
just one transition
$$ [0\;0\;1]\leftarrow 1,T_0\rightarrow [0\;0\;0] $$
between ferromagnetic and paramagnetic phases. For $U=U_3$, the transition
temperature $T_0$ corresponds to a tricritical point.

Finally, for $U_3<U<\infty$, we have the sequence
$$ [1\;1\;1]\leftarrow 2,T_1\rightarrow [0\;1\;1]\leftarrow 2,T_2\rightarrow
[0\;0\;1]\leftarrow 2,T_c\rightarrow[0\;0\;0] . $$

As is seen, the existence of heterophase mesoscopic fluctuations makes it
possible to get a rich variety of reorientation magnetic transitions which
are impossible in the pure case. Thus, mesoscopic fluctuations, at the
same time, provide global stability for a system and make its physics
much richer. Such fluctuations can also lead to specific features of
thermodynamic characteristics, related to pretransitional phenomena [1].
Probably, these fluctuations could be responsible for the existence of two
length scales at structural and magnetic phase transitions observed with
high resolution $X$--ray scattering techniques, particularly using
synchrotron sources [40].

\vspace{5mm}

{\bf Acknowledgement}

\vspace{2mm}

Financial support from the University of Western Ontario, Canada, is
appreciated.

\end{document}